# Molecular Communications with Longitudinal Carrier Waves: Baseband to Passband Modulation

Weisi Guo[1], Bin Li[2], Siyi Wang[3], Wei Liu[4],

***Abstract*—Traditional molecular communications via diffusion (MCvD) systems have used baseband modulation techniques by varying properties of molecular pulses such as the amplitude, the frequency of the transversal wave of the pulse, and the time delay between subsequent pulses. In this letter, we propose and implement passband modulation with molecules that exhibit longitudinal carrier wave properties. This is achieved through the oscillation of the transmitter. Frequency division multiplexing is employed to allow different molecular information streams to co-exist in the same space and time channel, creating an effective bandwidth for MCvD.**

## I. Introduction

Molecular communication has attracted significant research interest in recent years [1]. In terms of application, nano-robots in the medical domain will aim to track and operate on specific targets such as a tumor cell through sensing specific chemicals released by the cancerous region. Single robots of a few microns large that can perform specific simple tasks is already a reality [2]. However, communications between nano-robots in a nano-scale fluidic environment is needed, an environment that is hostile to the energy efficient generation and reliable propagation of acoustic and electromagnetic waves [3]. Molecular communications, which exists in nature at both the nano-scale and macro-scale [4], offers certain advantages over wave-based communication systems when there is a need for low energy transmission, as well as when there is an unacceptably high energy loss or input noise to wave propagation [5]. In such scenarios, traditional communication systems that rely on wave propagation may not achieve reliable communications. However, molecules that undergo random walk, which exhibit frequency-domain properties, may yet still assist with the delivery of information. This information is modulated onto the properties of molecules and together they can form an effective wireless information channel.

[1] Weisi Guo is with the School of Engineering, The University of Warwick, Coventry, CV4 7AL, UK. (E-mail: weisi.guo@warwick.ac.uk)

[2] Bin Li is with the School of Information and Communication Engineering (SICE), Beijing University of Posts and Telecommunications (BUPT), Beijing, 100876 China. (Email: stonebupt@gmail.com)

[3] Siyi Wang is with the Department of Electrical and Electronic Engineering, Xi'an Jiaotong-Liverpool University, Suzhou 215123, China. (E-mail: siyi.wang@xjtlu.edu.cn)

[4] Wei Liu is with the Communications Research Group, Department of Electronic and Electrical Engineering, The University of Sheffield, Sheffield, S1 3JD, UK. (Email: w.liu@sheffield.ac.uk)

This work of W. Guo has been supported by the University of Warwick's International Partnership Fund. This work of B. Li has been supported by Natural Science Foundation of China (NSFC) under Grants 61471061 and the Fundamental Research Funds for the Central Universities under Grant 2014RC0101. The work of S. Wang has been in part supported by the Research Development Fund (RDF-14-01-29) of Xi'an Jiaotong-Liverpool University.

### A. Review: Molecular Baseband Modulation

Fundamentally, molecular communications via diffusion (MCvD) involves modulating digital information onto the property of a single or a group of molecules. Regarding the diffusion channel, consider a 3-dimensional molecular diffusion channel with a transmitter and a receiver separated by distance $d$, with a molecular diffusivity $D$ and positive drift velocity $v$. The diffusion channel transfer function as a function of time $t$ is:

$$h(t) = \frac{1}{(4\pi Dt)^{\frac{3}{2}}} \exp\left[-\frac{(d-vt)^2}{4Dt}\right]. \quad (1)$$

For a fixed transmission distance of $d$, one can observe that there are essentially two main properties to modulate: the number of transmitter molecules $M$; and the pulse delay time $T_k$, such that the channel response is $h_k(t-T_k)$. For an input of binary symbols $a_k \in \mathcal{A} = \{0,1\}$, $k = 0, 1, \dots, \infty$, the output of the baseband modulator is $M_k$. Existing pulse modulation can be summarized as being one of the following:

- Amplitude or Concentration Shift Keying (ASK or CSK) [6], where the information is modulated into different levels of $M_k$, i.e., Binary ASK: $M_k \in \mathcal{M} = \{0, M\}$.
- Frequency shift keying (FSK) [7], where a sinusoidal pulse of a variable frequency $f$ is emitted $M_k(f_k) = M\sin(2\pi f_k t)$, i.e., Binary FSK: $f_k \in \mathcal{F} = \{0, f\}$.
- Pulse Position Modulation (PPM) [8], where the information is modulated into the bit delay time $T$, i.e., Binary PPM: $T_k \in \mathcal{T} = \{0, T\}$.

All of the aforementioned modulation schemes has been successfully implemented in hardware, achieving reliable data transfer over a few metres of both free space [9] and maze environments [10]. In addition, the chemical type can be used to encode information, as it is common in nature, which is known as Molecule Shift Keying (MoSK) [11]. However, the complexity of synthesizing and detecting even a small number ($\sim 10$) of chemical compositions is complex, expensive, and remains largely theoretical [12].

### B. Contribution: Molecular Passband Modulation

The aforementioned modulations can be regarded as *baseband modulation*, whereby the resolution of the discrete modulation constellations is fundamentally limited by the inter-symbol-interference (ISI) and the stochastic behaviour of diffusion. Whilst significant efforts have been made towards reducing ISI through coding and signal processing means [13], [14], baseband MCvD communication can only achieve a limited data rate [8], typically less than 1 bit/s per chemical type [9]. Without an achievable bandwidth, it is difficult to upscale the data rate of molecular communications. The concept of a carrier wave, such as that associated with our pre-existing knowledge of electromagnetic (EM) wave communications, has been missing in MCvD. This is due to the fact that until now, MCvD systems lack a continuous wave concept in what is fundamentally a discrete Gaussian kernel diffusion model.

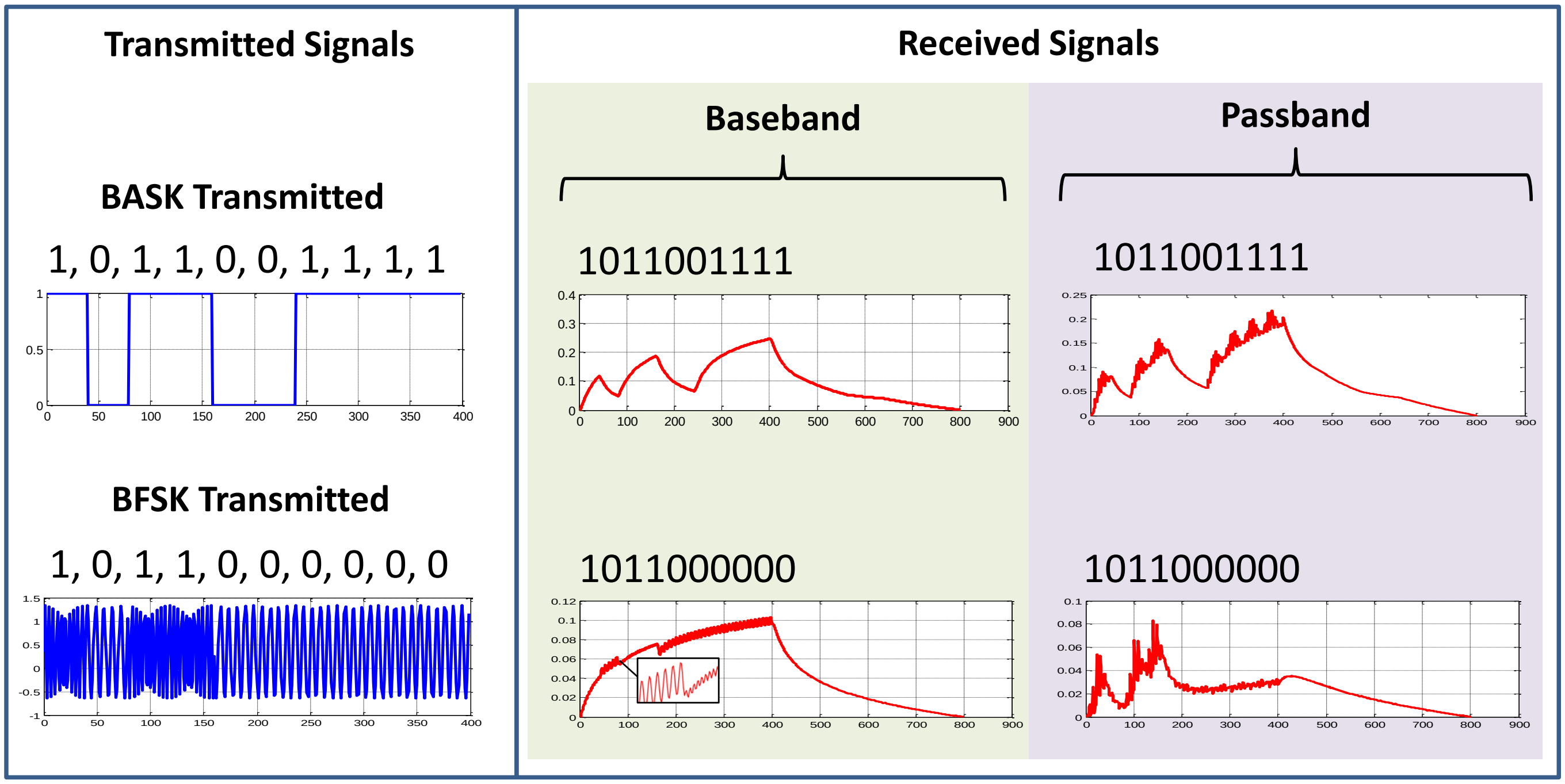

Fig. 2. Illustration of two different MCvD modulation techniques and the resulting baseband and passband received signals.

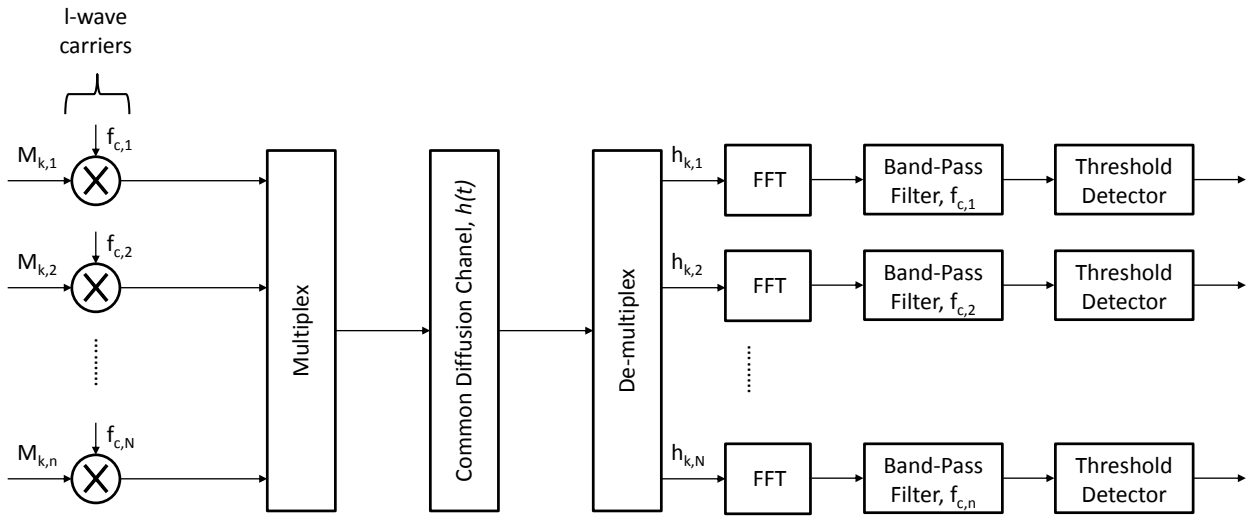

Fig. 1. Illustration of the modulation, multiplexing, de-multiplexing and demodulation process.

This letter sets out to introduce how longitudinal-waves can be added to the aforementioned baseband modulation schemes to create a multiple access channel. We show that multiple independent data streams can be multiplexed and de-multiplexed together.

## II. Carrier Signal: Longitudinal-Waves

In most practical molecular diffusion systems [9], real time communications is achieved with an initial release velocity $v$ from the chemical emitter. In effect this creates molecules that exhibit a *longitudinal compression wave* (l-wave) property, which can be considered as a carrier wave, one that is independent of the aforementioned baseband modulation schemes (i.e., amplitude, pulse delay, and transversal frequency of pulses). By being able to control different longitudinal carrier wave frequencies $f_{c,n}$, there is *potential* for a limitless number of orthogonal molecular communication channels.

Let us now consider $n \leq N$ unique information streams, each transmitting using the same molecular compound and using the same baseband modulation technique. They share a common diffusion channel $h(t)$. In order to reliably multiplex and de-multiplex $N$ signals, we propose a carrier frequency concept. As shown in Fig. 1, each baseband signal $M_{k,n}$ will be modified by a carrier frequency of $f_{c,n}$.

### A. Oscillating Transmitter

EM carrier signals are transverse waves, where the oscillations occur perpendicular to the direction wave travels. In molecular communications, the wave generated by the movement of particles is longitudinal. Consider a transmitter and a receiver that is separated by a distance $d_0$ and the transmitter is allowed to oscillate such that the instantaneous transmission distance varies according to:

$$d(t) = d_0 + A_c \sin(2\pi f_{c,k} t), \tag{2}$$

where $A_c$ is the peak amplitude of oscillation and $f_{c,k}$ is the frequency of the carrier signal.

Then, the channel response $y_{k,n}(t)$ derived from Eq.(1) is:

$$y_{k,n}(t) = \frac{M_{k,n}}{(4\pi D t)^{\frac{3}{2}}} \exp\left[ -\frac{(d_0 + A_c \sin(2\pi f_{c,k} t) - vt)^2}{4Dt} \right]. \tag{3}$$

In Fig. 2, we show the baseband and passband results for a binary ASK/CSK modulation and a binary FSK modulation scheme. In this time domain representation, it can be seen that an oscillatory component has been added, as well as non-linear effects due to the exponential term in the diffusion channel model given in Eq.(1).

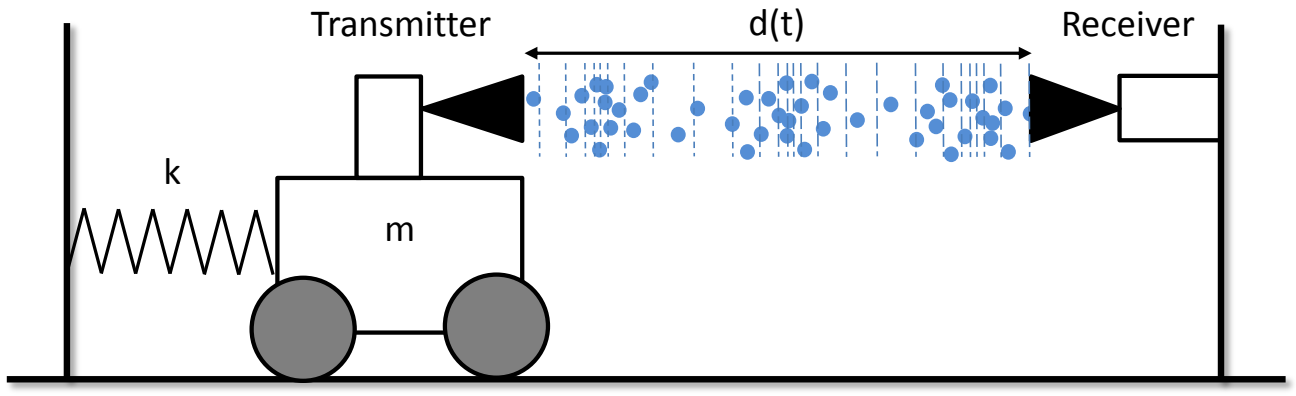

Fig. 3. Illustration of a potential l-wave carrier signal generation method using an oscillating transmitter.

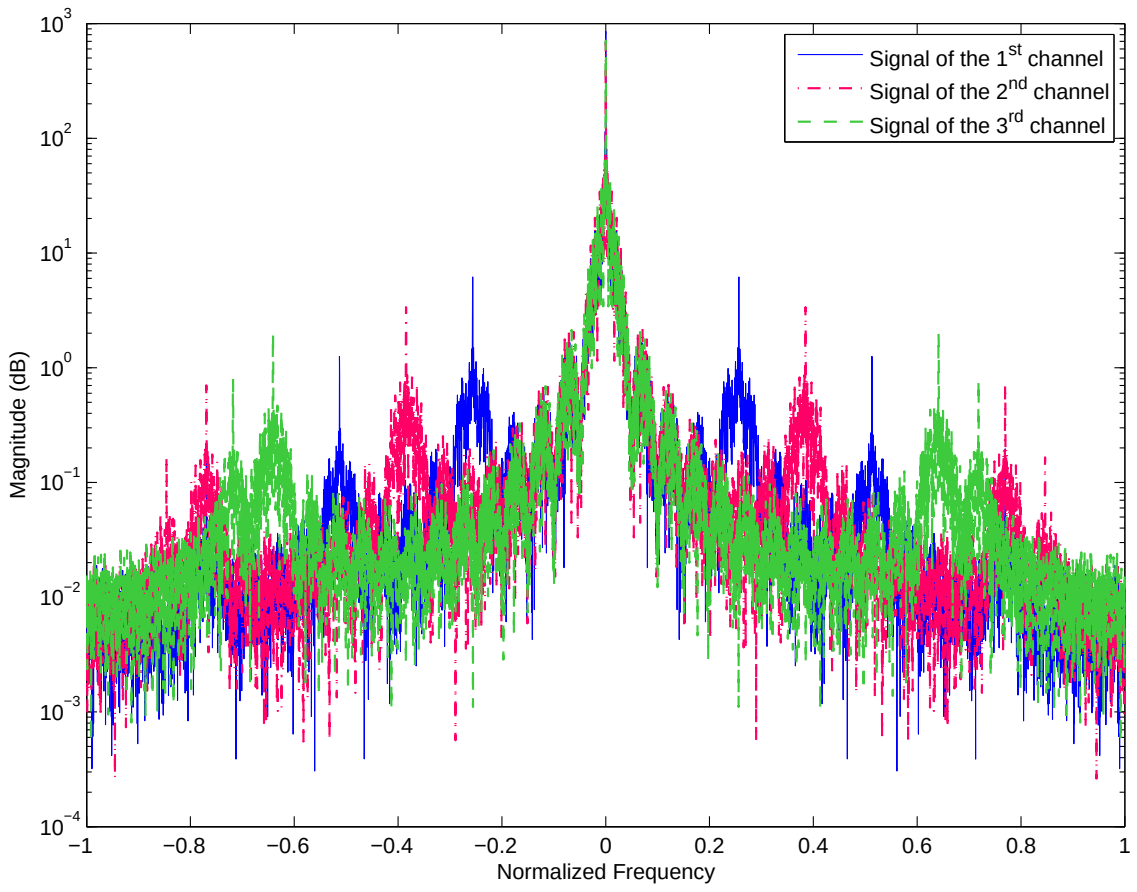

Fig. 4. Frequency response of $N = 3$ molecular signals multiplexed over a single diffusion channel.

## *B. Multiplexing and De-multiplexing*

In terms of multiplexing implementation, one way this can be implemented is by attaching a spring of stiffness $K$ (N/m) to each transmitter (mass $m$), such that the l-wave carrier frequency is given by: $f_c = \frac{1}{2\pi}\sqrt{\frac{K}{m}}$. This is illustrated in Fig. 3 for a well understood example of creating l-waves, where the spring stiffness can be adjusted to create different carrier frequencies. Alternative implementations are less mechanical, and can involve digitally controlled compression wave generators at the transmitter.

As mentioned previously, the time domain output in Fig. 2 demonstrates non-linear effects of the diffusion channel. We now consider $N = 3$ independent data channels multiplexed together, each with the same baseband BASK modulation. At the common receiver, after FFT, the frequency response can be seen in Fig. 4. It can be seen that the baseband signal is at 0 normalized frequency. The first harmonic of each signal can be seen distinctively. To view the frequency response more clearly, we extract the $n = 1$ signal in Fig. 5(top) with its harmonic peaks. The first peak is the useful signal.

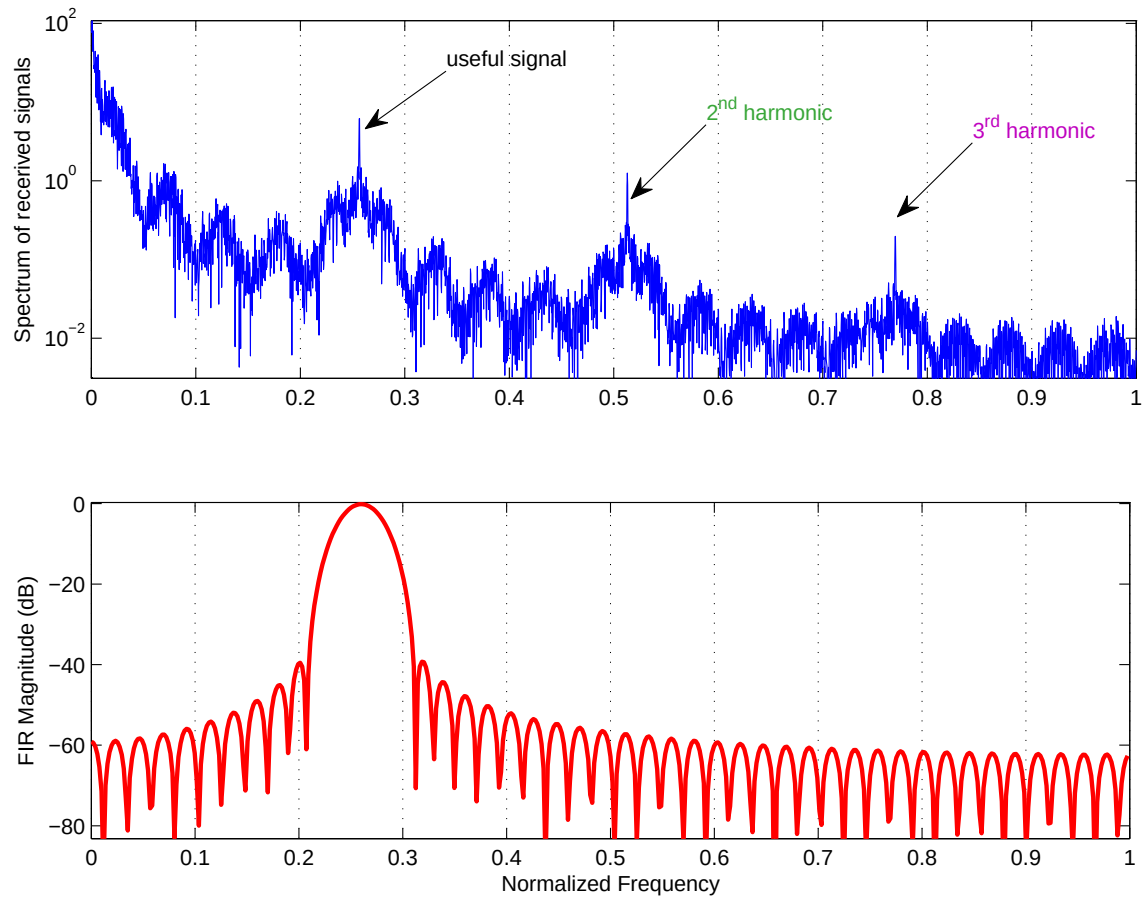

Fig. 5. (top) Frequency response of $n = 1$ molecular signal with its harmonic peaks; (bottom) FIR magnitude of the bandpass filter designed to filter the first harmonic (useful signal).

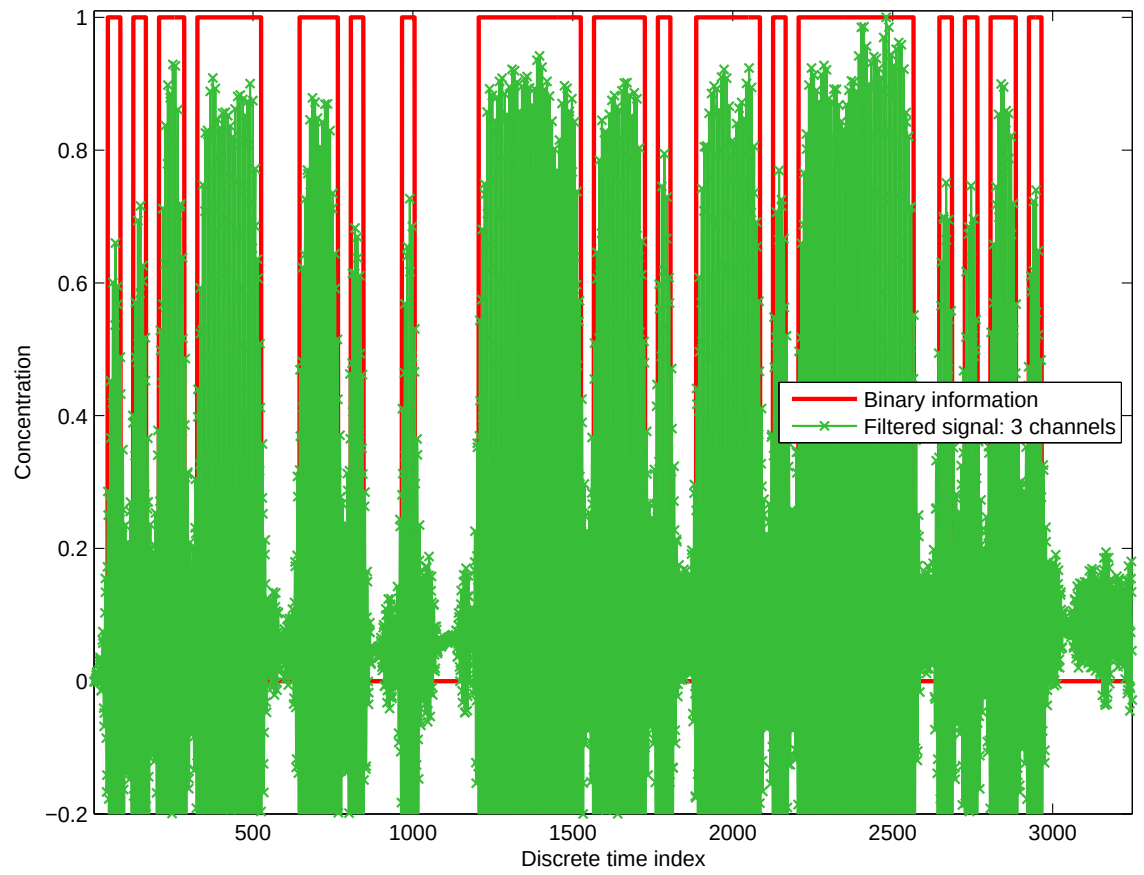

Fig. 6. Post-filtering results for a order 90 Kaiser bandpass filter.

Fig. 5(bottom) shows the FIR magnitude of the bandpass filter designed to filter the useful signal. We use a Kaiser filter (parameter 3.3953) with a centre frequency of $f_{c,n}$ for each $n$ signal and a transition bandwidth of 0.05 normalized frequency. Due to the narrow nature of the bandpass filter, the FIR filter order is 86.

Fig. 6 shows the post-filtering results in the time domain for $n = 1$ in $N = 3$ multiplexed signals. The results show that the original modulated signal can be fully reconstructed. A similar set of results can also be found for using BFSK baseband modulation using the same methodology.

The parameters for the simulated results are as follows: distance $d_0 = 5$m for all channels, diffusivity $D = 1$cm$^2$/s, the amplitude of oscillation is $A_c = 1.67$m, and the carrier frequencies for the $N = 3$ links are: $f_{c,1} = 5$Hz, $f_{c,2} = 7.5$Hz, and $f_{c,3} = 12.5$Hz.

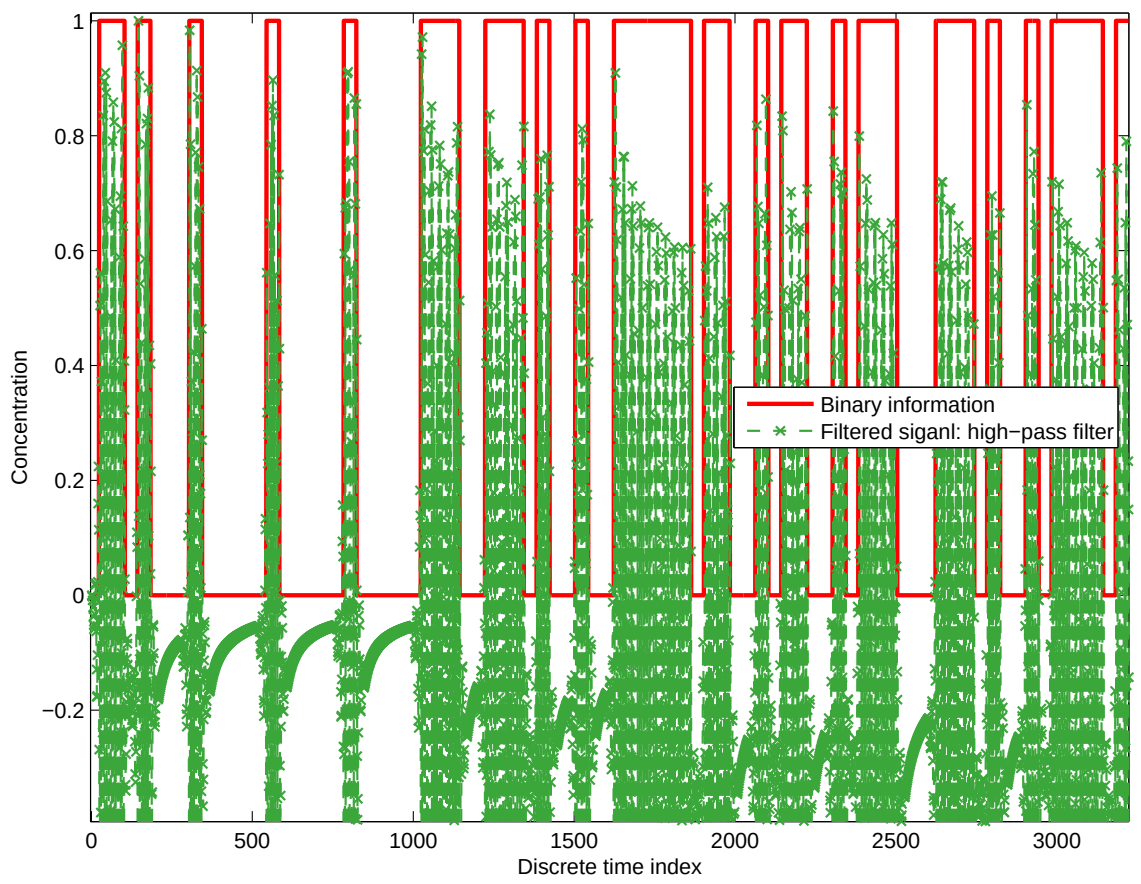

Fig. 7. Post-filtering results for a order 47 Kaiser highpass filter.

## III. Discussion and Future Work

One promising application of molecular communications is likely to be in the nano-scale dimension, especially enabling communication between nano-robots [3]. In order to build energy efficient transmitter and receiver circuits, one particular challenge faced at the receiver side is the need for low-order narrow band bandpass filters, particularly those that can deal with the non-linear effects of the channel. The aforementioned FIR Kaiser bandpass filter needs a FIR filter order of 86. This complexity can be further reduced by using a Kaiser highpass filter, which has a filter order 47. The result is that the sidebands will be included in the filtered signal, but it appears that due to the decaying nature of the sidebands, this has a negligible effect on the final de-multiplexed signal, as shown in Fig. 7. There is scope here for further investigation to find appropriate filter designs for optimal de-multiplexing of molecular signals, with special consideration to the non-linear nature of the channel and the complexity of the filters.

Another area of challenge is transmit peak chemical concentration control (analogy to uplink power control). This paper has assumed that the transmission distance for all the channels is fixed and the same. In reality, a multiple access channel is likely to be shared with transmitters at different distances and act as co-chemical channel interference to each other. This is due to the sidebands exhibited in the passband signals shown in Fig.4. The sidebands can not be suppressed due to the diffusion nature of the channel, and any transmit pulse shapes will be diluted in the diffusion process. The sidebands of one channel effectively act as interference for other channels. In order to achieve a similar signal-to-interference ratio (SIR) to each other, careful transmit concentration control must be utilized.

## IV. Conclusion

In this paper we have presented a viable way of scaling the data rate of molecular communications by combining baseband modulation techniques with a longitudinal carrier wave generated by an oscillating transmitter. This to the best of our knowledge is the first proven method to create bandwidth for multiple access molecular communications. Our results have shown that $N$ independent data streams using a common baseband modulation technique such as ASK or FSK can be multiplexed together using different carrier waves and then reliably de-multiplexed using bandpass or highpass filters. The authors also point towards two promising areas of research for future work on communication between nano-robots, namely: low-complexity energy efficient filters suited towards non-linear molecular signals, and transmit concentration control for mobile molecular communications.